\documentclass[superscriptaddress,twocolumn]{revtex4-2}
\usepackage{graphicx}
\usepackage{multirow}
\usepackage{placeins}
\usepackage{dcolumn}
\usepackage{bm}
\usepackage[colorlinks=true, citecolor=cyan, urlcolor = violet, linkcolor= purple, bookmarks=true]{hyperref}
\usepackage{float}
\usepackage{amsmath,amsfonts,amssymb}
\usepackage{natbib}
\usepackage{hyperref}
\usepackage{breakurl}
\usepackage{xcolor,verbatim,multirow}
\usepackage[normalem]{ulem}
\usepackage{subcaption} 
\usepackage{wrapfig}
\usepackage{adjustbox}
\usepackage{caption}
\begin{document}

\title{Testing EGB gravity coupled to bumblebee field and black hole parameter estimation with EHT observations}

\author{Misba Afrin}
\email{me.misba@gmail.com}
\affiliation{Saha Institute of Nuclear Physics, 1/AF Bidhannagar, Kolkata 700064, India}
\author{Sushant G. Ghosh} \email{sghosh2@jmi.ac.in}
\affiliation{Centre for Theoretical Physics, Jamia Millia Islamia, New Delhi 110025, India}
\affiliation{Astrophysics and Cosmology	Research Unit, School of Mathematics, Statistics and Computer Science, University of KwaZulu-Natal, Private Bag 54001, Durban 4000, South Africa}
\author{Anzhong Wang}
\email{anzhongwang@baylor.edu}
\affiliation{GCAP-CASPER, Physics Department, Baylor University, Waco, Texas 76798-7316, USA}
\begin{abstract}
A general covariant Einstein--Gauss--Bonnet Gravity in Four-Dimensional (4D EGB) spacetime is shown to bypass Lovelock's theorem and is free from Ostrogradsky instability. Meanwhile, the bumblebee theory is a vector--tensor theory. It extends the Einstein--Maxwell theory that allows for the spontaneous symmetry breaking that leads to the field acquiring a vacuum expectation value, introducing Lorentz violation into the system. We investigate rotating black holes in the 4D EGB-bumblebee gravity model where Lorentz symmetry is spontaneously broken -- Kerr EGB bumblebee (KEGBB) black holes. The latest observations from the \textit{Event Horizon Telescope} (\textit{EHT}) of the supermassive black holes (SMBHs) M87* and Sgr A* have sparked intensified interest in the study of black hole shadows, which present a novel avenue for investigating SMBHs within the strong-field regime. Motivated by this, we model SMBHs M87* and Sgr A* as KEGBB black holes, and using the EHT observation result, for given $l$, to find earlier upper limits on the $\alpha$ and $a$ are altered. The KEGBB and Kerr black holes are indiscernible in some parameter space, and one cannot rule out the possibility that the former may serve as strong candidates for astrophysical black holes. Employing our newly developed parameter estimation technique, we use two EHT observables -- namely, the angular diameter of the shadow, $d_{sh}$, and the axial ratio, $\mathcal{D}_A$ -- to estimate parameters of M87* and Sgr A* taking into account observational errors associated with the EHT results.

\end{abstract}
\pacs{}
                              
\maketitle

\section{Introduction}\label{Intro}
General Relativity (GR) remains the most extensively tested gravity model on a macroscopic level. Despite this, since its inception, many theoretical ventures have sought extensions to GR to formulate more unified theories that incorporate extra fields \citep{Brans:1961sx,Horndeski:1974wa,1975GReGr6259R,Jacobson:2007veq}, higher derivative theories \citep{Buchdahl:1970ynr,Starobinsky:1980te,Nicolis:2008in,Horava:2009uw}, and theories in higher dimensions \citep{Kaluza:1921tu,Randall:1999ee,Dvali:2000hr,Lovelock:1971yv}. One of the most comprehensive alternative gravity theories in higher dimensions is the Einstein-Gauss-Bonnet (EGB) gravity \citep{Lovelock:1971yv}.  String theory's low energy limits suggest that the gravity action should consist of quadratic and higher-order curvature terms in dimensionally continued GB densities. If only quadratic curvature GB terms are included, the resulting theory is the EGB gravity, which is valid only in $D>4$ 
 dimensions \citep{Lanczos:1938sf}.  The GB term is a quadratic correction term that involves the contraction of the Riemann tensor and can be viewed as a topological invariant in four dimensions ($4D$) \citep{Kumar:2020owy}. However, in higher dimensions, this term can have a nontrivial contribution to the dynamics of the theory.
Boulware and Deser \cite{Boulware:1985wk} first discovered exact black hole solutions in $D(\geq 5)$-dimensional EGB gravitational theory. Subsequently, the charged counterpart has been also obtained \citep{Wheeler:1985nh,Wiltshire:1985us}. In $4D$ spacetime, the Gauss-Bonnet term becomes a total derivative and does not contribute to the gravitational dynamics \citep{Lanczos:1938sf,1987NuPhB.291...41G,Boulware:1985wk,Bento:1995qc,Wheeler:1985nh,Wiltshire:1985us}.  Glavan and Lin \cite{Glavan:2019inb} proposed a novel regularised 4D EGB gravity and achieved this by rescaling the GB coupling with $\alpha/(D-4)$ and then taking the equations of motion's limit at $D = 4$, which resulted in significant contributions from the GB term in the 4D spacetime equations of motion. Tomozawa \cite{Tomozawa:2011gp} originally suggested this regularization procedure to account for finite one-loop quantum corrections to Einstein gravity. Tomozawa's  \citep{Tomozawa:2011gp} approach predicted a spherically symmetric black hole solution with a repulsive gravity nature at short distances. Cognola  {\it et al.} \cite{Cognola:2013fva} simplified Tomozawa's \citep{Tomozawa:2011gp} approach by reformulating the arguments that imitate quantum corrections due to a GB invariant within a classical Lagrangian approach.  Furthermore, Glavan \& Lin \cite{Glavan:2019inb} have shown that the theory contains the degrees of freedom only of massless graviton as in GR and thus free from the ghosts \citep{Glavan:2019inb}.  Hence, the  $4D$ EGB gravity theory drew considerable attention \citep{Konoplya:2020bxa,Guo:2020zmf,Fernandes:2020rpa,Wei:2020ght,Doneva:2020ped,Ghosh:2020vpc,Ghosh:2020ijh,Kumar:2020sag,Islam:2020xmy,Ghosh:2020syx,Ghosh:2020cob,Kumar:2020xvu,Bazeia:2023czl}.

 The EGB gravity theory can be coupled to a vector field, known as the bumblebee field \citep{Kostelecky:2003fs,Maluf:2020kgf,Maluf:2014dpa,Heidari:2024bvd}, which couples to the Ricci tensor $R_{\mu\nu}$ in a way similar to the electromagnetic four-potential in electromagnetism \citep{Kostelecky:2003fs}, the resulting theory is called EGB-bumblebee gravity model. The bumblebee gravity model has been extensively studied in the weak-field regime  \citep{Kostelecky:2003fs,Bluhm:2004ep,Bluhm:2005uj,Bailey:2006fd,Bluhm:2007bd,Bluhm:2008yt,Liang:2022hxd} and more recently an exact vacuum solution within metric affine bumblebee gravity has been obtained \citep{Filho:2023etf,Filho:2022yrk,AraujoFilho:2024ykw}. The original motivation for the bumblebee formalism was based on ideas from string theory, which suggest that tensor-valued fields can acquire vacuum expectation values and result in the spontaneous breaking of Lorentz symmetry \cite{Maluf:2013nva}. In this context, the bumblebee field can attain a background configuration that minimizes the potential $V$. As a result, it can have a nontrivial background configuration, leading to significant modifications in the qualitative features of black holes in the theory, as demonstrated by Kosteleck\'y \cite{Kostelecky:2003fs}. 
 
Adding the bumblebee field to EGB gravity introduces a new degree of freedom that can affect the behaviour of black holes in the theory. The spherically symmetric static black hole solution in this theory has been studied \citep{Ding:2021iwv}. Further, it has been shown that the bumblebee field can have a significant impact on the horizon structure of the black hole, as well as on its thermodynamic properties, viz., it can cause the formation of a new horizon outside the event horizon of the black hole, known as the bumblebee horizon, which can have a nontrivial thermodynamic entropy \citep{Karmakar:2023mhs,Mai:2023ggs}. The bumblebee field has been shown to affect the shape, size, asymmetry, and distortion of the shadow, providing a way to test the theory against observations of black holes via their shadows \citep{Wei:2019pjf,Chen:2019fsq,Xu:2023xqh,Jha:2020pvk}; the effects in gravitational lensing have also been studied \cite{Filho:2024isd,Filho:2023etf}. The study of the EGB-bumblebee gravity model can shed light on several conceptual issues of gravity in a broader setup and help us understand the role of additional fields in modifying the behaviour of black holes. The theory has also been studied in gravitational waves and cosmology \citep{Neves:2022qyb} and has potential implications for detecting gravitational waves from black hole mergers \cite{Amarilo:2023wpn} and quasinormal modes \cite{Oliveira:2018oha,Filho:2023etf}.

To test the model and Lorentz symmetry in the strong-field regime, obtaining strong-field solutions is increasingly important to use astrophysical observations of black hole shadows \citep{EventHorizonTelescope:2019dse,EventHorizonTelescope:2022exc}. In this regard, we take the first step by investigating Kerr black holes in the EGB-bumblebee model and using the shadows of the two supermassive black holes (SMBHs) captured by the Event Horizon Telescope (EHT) collaboration to limit the bumblebee field's behaviour around the black holes.
Furthermore, we investigate the horizons and shadows that these black holes cast. We use EHT shadow observables to characterize the shadow's shape, size, and distortion and estimate the black hole parameters. The scenario in this paper is that we consider M87* and Sgr A* as Kerr black holes in the EGB-bumblebee model and use the results of the EHT. Our results put simultaneous bounds on the coupling constant $\alpha$ in the presence of the bumblebee parameter $l$ of the black holes for the first time.
\section{General Model}

The bumblebee theory is a vector-tensor theory and an extension of the Einstein-Maxwell theory. The action of EGB gravity coupled to the bumblebee vector field $B_{\mu}$, in $D-$dimensional spacetime, reads \citep{Ding:2021iwv},
\begin{align}
\mathcal{S}=&
\int d^Dx\sqrt{-g}\Big[\frac{R}{2\kappa}+\frac{2\alpha}{D-4} \mathcal{G}+\frac{\Omega}{2\kappa} B^{\mu}B^{\nu}R_{\mu\nu}\nonumber\\
&-\frac{1}{4}B^{\mu\nu}B_{\mu\nu}
-V(B_\mu B^{\mu}\mp b^2)+\mathcal{L}_M\Big], \label{action}
\end{align}
where  $\kappa=8\pi G$, $R$ is Ricci scalar,  and
$\alpha$ is the GB coupling parameter and the GB invariant $\mathcal{G}$ has the form \citep{Boulware:1985wk}
\begin{align}\label{}
\mathcal{G}=R_{\mu\nu\tau\sigma}R^{\mu\nu\tau\sigma}-4R_{\mu\nu}R^{\mu\nu}+R^2,
\end{align}
The coupling constant $\Omega$ indicates the strength of quadratic interaction with bumblebee field $B_{\mu\nu}$ and Ricci tensor $R_{\mu\nu}$. The field strength of the bumblebee field is $B_{\mu\nu}=\partial_{\mu}B_{\nu}-\partial_{\nu}B_{\mu}$, due to antisymmetry of $B_{\mu\nu}$. The term $\mathcal{L}_M$ represents possible interactions with matter or external currents.
The constant $b$ is a real positive constant. The  potential $V(B_\mu B^{\mu}\mp b^2)$ has a minimum at $B^{\mu}B_{\mu}\pm b^2=0$ and $V'(b_{\mu}b^{\mu})=0$ to ensure the destroying of the $U(1)$ symmetry, where the field $B_{\mu}$ acquires a nonzero VEV, $\langle B^{\mu}\rangle= b^{\mu}$.
The minimum potential will ensure a stable vacuum of spacetime \citep{Mai:2023ggs,Ding:2019mal}. Varying the action (\ref{action}) with respect to the $g_{\mu\nu}$ we obtain the gravitational field equations
\begin{align}\label{einstein0}
G_{\mu\nu}=R_{\mu\nu}-\frac{1}{2}g_{\mu\nu}R=\kappa T_{\mu\nu}^B+2\alpha\kappa T^{GB}_{\mu\nu}+\kappa T_{\mu\nu}^M,
\end{align}
where the momentum tensor $T_{\mu\nu}^B$ due to bumblebee energy reads \citep{Ovgun:2018xys}
\begin{align}\label{momentum}
T_{\mu\nu}^B=&-B_{\mu\alpha}B^{\alpha}_{\;\nu}-\frac{1}{4}g_{\mu\nu} B^{\alpha\beta}B_{\alpha\beta}- g_{\mu\nu}V+
2B_{\mu}B_{\nu}V'\nonumber\\
+&\frac{\Omega}{\kappa}\Big[\frac{1}{2}g_{\mu\nu}B^{\alpha}B^{\beta}R_{\alpha\beta}
-B_{\mu}B^{\alpha}R_{\alpha\nu}-B_{\nu}B^{\alpha}R_{\alpha\mu}\nonumber\\
+&\frac{1}{2}\nabla_{\alpha}\nabla_{\mu}(B^{\alpha}B_{\nu})-\frac{1}{2}
g_{\mu\nu}\nabla_{\alpha}\nabla_{\beta}(B^{\alpha}B^{\beta})
\nonumber\\
-&\frac{1}{2}\nabla^2(B_{\mu}B_{\nu})\Big],
\end{align}
whereas, the GB energy momentum tensor $T^{GB}_{\mu\nu}$ is \citep{Boulware:1985wk},
\begin{align}\label{momentumG}
T_{\mu\nu}^{GB}=&4R_{\alpha\beta}R^{\alpha\;\beta}_{\;\mu\;\nu}-2R_{\mu\alpha\beta\gamma}
R_{\nu}^{\;\alpha\beta\gamma}+4R_{\mu\alpha}R^{\alpha}_{\;\nu}\nonumber\\
-&2RR_{\mu\nu}
+\frac{1}{2}g_{\mu\nu}\mathcal{G}.
\end{align}

To obtain a black hole solution, we suppose that there is no matter field and the bumblebee field is frosted at its VEV  \citep{Casana:2017jkc,Bertolami:2005bh}, i.e., $B_\mu=b_\mu$
The particular form of the potential that governs the system's dynamics is insignificant, resulting in $V=0,\; V'=0$. Consequently, the first two terms in Eq. (\ref{momentum}) resemble those of the electromagnetic field, with the only distinguishing feature being the coupling terms to the Ricci tensor. When this condition is met, Eq. (\ref{einstein0}) gives rise to the gravitational field equations 
\begin{align}\label{bar}
G_{\mu\nu}=&2\alpha\kappa T^{GB}_{\mu\nu}+\kappa (b_{\mu\alpha}b^{\alpha}_{\;\nu}-\frac{1}{4}g_{\mu\nu} b^{\alpha\beta}b_{\alpha\beta})+\bar B_{\mu\nu}\nonumber\\
+&\Omega\Big(\frac{1}{2}
g_{\mu\nu}b^{\alpha}b^{\beta}R_{\alpha\beta}- b_{\mu}b^{\alpha}R_{\alpha\nu}
-b_{\nu}b^{\alpha}R_{\alpha\mu}\Big),
\end{align}
with
\begin{align}\label{barb}
\bar B_{\mu\nu}=&\frac{\Omega}{2}\Big[
\nabla_{\alpha}\nabla_{\mu}(b^{\alpha}b_{\nu})
+\nabla_{\alpha}\nabla_{\nu}(b^{\alpha}b_{\mu})
-\nabla^2(b_{\mu}b_{\nu})\nonumber\\
-&g_{\mu\nu}\nabla_\alpha\nabla_\beta(b^\alpha b^\beta)\Big].
\end{align}

We now derive the static black hole solution of the 4D EGB-bumblebee gravity given by action \eqref{action}. Taking the $D$-dimensional static and spherically symmetric metric and solving the field equation (\ref{bar}) in the limit $D \to 4$ \citep{Ghosh:2020vpc,Ghosh:2020syx}, we get the static spherically symmetric black hole solutions in the EGB-bumblebee gravity as \cite{Ding:2021iwv}
\begin{align}\label{metric_spherical}
ds^2=- (1+l)f(r)dT^2+\frac{1}{f(r)}dr^2+r^2d\theta^2
+r^2\sin^2\theta d\phi^2,
\end{align}
with
\begin{align}\label{f_r}
f(r)=1+\frac{(1+l)r^2}{32\pi\alpha G}\left(1\mp\sqrt{1+\frac{64\pi\alpha lG^2}{(1+l)^2r^2}
+\frac{128\pi\alpha MG^2}{(1+l)^2r^3}}\right).
\end{align}
Here, $l=\Omega b^2$ is the Lorentz violation parameter. If $l\rightarrow0$, Eq.~\eqref{f_r} recovers the $4D$ EGB black hole metric \citep{Glavan:2019inb}. There are two branches of solutions for the metric function (\ref{f_r}) with $\alpha>0$, the first one corresponding to the $``-"$ sign and the second to $``+"$ sign.
When $\alpha\rightarrow0$, the $-$ve branch leads to Schwarzschild-like black holes 
\citep{Casana:2017jkc}, and Boulware and Deser \cite{Boulware:1985wk} have shown that the $+$ve branch is unstable and leads to graviton ghosts.
When $r\rightarrow\infty$, the two branches have similar behave asymptotically at large distances, i.e., the first is asymptotic to a Schwarzschild-like black hole with positive mass $M$, the second to a Schwarzschild-de Sitter like black hole with negative mass $M$.
In our study, we shall focus on the first solution, i.e.black, the $-$ve branch of the metric (\ref{f_r}) and let $16\pi G=1$. It is easy to see that their two horizons locate at $g_{00}(r_\pm)=0$,
$r_\pm^H=M\pm\sqrt{M^2-\alpha}$; clearly, the parameter $l$ doesn't influence the horizons of the spherically symmetric black holes \eqref{metric_spherical} with \eqref{f_r}. If the mass $M<M_*=\sqrt{\alpha}$ is the critical mass, then there is no horizon and consequently no black hole solutions. 

\subsection{Rotating black hole metric using  modified Newman-Janis algorithm}

Next, we shall consider the rotating counterpart of the metric \eqref{metric_spherical}. It will be useful to redefine the time coordinate $T$ as $$T=\frac{t}{\sqrt{1+l}},$$ with which the metric \eqref{metric_spherical} simplifies to 
\begin{align}\label{metric_spherical_transformed}
ds^2=- f(r)dt^2+\frac{1}{f(r)}dr^2+r^2 (d\theta^2
+\sin^2\theta d\phi^2).
\end{align}
The Newman-Janis algorithm is widely used to construct rotating black hole solutions from their spherically symmetric counterparts \citep{Newman:1965tw}. We use the non-complexification procedure \citep{Azreg-Ainou:2014aqa, Azreg-Ainou:2014pra} or the modified Newman-Janis algorithm --- which modifies the Newman-Janis algorithm and has been effective in generating rotating metrics for imperfect fluids from static spherically symmetric metrics \citep{Newman:1965tw,Johannsen:2011dh,Jusufi:2019caq,Bambi:2013ufa,Ghosh:2014hea,Ghosh:2015ovj,Ghosh:2014pba,Ghosh:2021clx,Walia:2021emv} --- to construct stationary or rotating spacetimes based on the static seed metric  (\ref{metric_spherical_transformed}).

The first step is to express the metric mentioned above in the advanced null (Eddington-Finkelstein) coordinates $(u, r, \theta, \phi)$; for this, we define the transformation
\begin{equation}
	du=dt-\frac{dr}{f(r)},
\end{equation}
the static seed  metric (\ref{metric_spherical_transformed}) in the advance null coordinate takes the form 
\begin{equation}
	ds^2=-f(r) du^2-2dudr+r^2\left(d\theta^2+\sin^2\theta d\phi^2\right).
\end{equation}
We introduce the set of null tetrad $Z_\alpha^\mu=(l^\mu,n^\mu,m^\mu,\bar{m}^\mu)$, such that the inverse metric $g^{\mu\nu}$ are of the form 
\begin{equation}
	g^{\mu\nu}=-l^\mu n^\nu -l^\nu n^\mu +m^\mu \bar{m}^\nu +m^\nu \bar{m}^\mu,\label{g1}
\end{equation}
with
\begin{equation}
	l^\mu=\delta^\mu_r, \;\; n^\mu=\delta^\mu_u-\frac{f(r)}{2}\delta^\mu_r, \;\; m^\mu=\frac{1}{\sqrt{2}r}\left(\delta^\mu_\theta+\frac{i}{\sin\theta}\delta^\mu_\phi\right).
\end{equation}
and $\bar{m}^\mu$ is the complex conjugate of $m^\mu$. The null tetrad vectors are  orthonormal and satisfy the following relations
\begin{equation}
	l_\mu l^\mu = n_\mu n^\mu = m_\mu m^\mu = l_\mu m^\mu = n_\mu m^\mu =0,
\end{equation}
and also
\begin{equation}
	l_\mu n^\mu = - m_\mu \bar{m}^\mu =-1.
\end{equation}
Next, we execute the complex coordinate transformation wherein the $\delta^{\mu}_{\nu}$ transfromed as vectors \citep{Azreg-Ainou:2014pra}
\begin{equation}
	\delta^{\mu}_{r}\to \delta^{\mu}_{r},\;\; \delta^{\mu}_{u}\to \delta^{\mu}_{u},\;\; \delta^{\mu}_{\theta}\to \delta^{\mu}_{r} +ia\sin\theta(\delta^{\mu}_{u}-\delta^{\mu}_{r}),\;\; \delta^{\mu}_{\phi}\to \delta^{\mu}_{\phi}.
\end{equation}
Here, $a$ represents the spin parameter of the black hole. The revised Newman-Janis algorithm by Azreg-Ainou  \citep{Azreg-Ainou:2014pra}, eliminates the ambiguity associated with the complexification of the radial coordinate. Instead, the algorithm treats function $f(r)$ as transforming into $F(r, a, \theta)$ and $r^2$ as transforming into $H(r, a, \theta)$. Following procedure ~\citep{Azreg-Ainou:2014pra}, we find the transformed null tetrads read as 
\begin{align}
	l'^\mu=&\delta^\mu_r, \quad n'^\mu=\delta^\mu_u-\frac{F(r,a,\theta)}{2}\delta^\mu_r,\nonumber\\
	m'^\mu=&\frac{1}{\sqrt{2H(r,a,\theta)}}\left(ia\sin\theta(\delta^\mu_u-\delta^\mu_r)+\delta^\mu_\theta+\frac{i}{\sin\theta}\delta^\mu_\phi\right).\label{t1}
\end{align}
This transformed tetrad yields a new
metric.  Using eq.~(\ref{t1}) in definition (\ref{g1}), we obtain the new inverse metric as 
\begin{equation}
	g^{\mu\nu}=-l'^\mu n'^\nu -l'^\nu n'^\mu +m'^\mu \bar{m}'^\nu +m'^\nu \bar{m}'^\mu,
\end{equation}
which gives the rotating black hole metric in the Eddington-Finkelstein coordinates as follows
\begin{eqnarray}\notag
	ds^2&=&-F(r,a,\theta)du^2-2dudr+2a\sin^2\theta\left(F(r,a,\theta)-1\right)du d\phi\nonumber\\
      &&+ 2a\sin^2\theta drd\phi+H(r,a,\theta) d\theta^2+\sin^2\theta\Big[H(r,a,\theta)\nonumber\\
	&&+a^2\sin^2\theta\left(2-F(r,a,\theta)\right)\Big]d\phi^2.
	\label{g3}
\end{eqnarray}
Here $F(r, a,\theta)$ is a function which depends on $f(r)$. We have applied the aforesaid procedure to spherical metric (\ref{metric_spherical_transformed})
However, the method is robust and suitably applies to any spherically symmetric solution to generate rotating spacetimes \citep{Kumar:2022fqo, Afrin:2021wlj, Walia:2021emv, Kumar:2020owy, Brahma:2020eos, Islam:2022wck}.   
The last yet crucial step involves converting the metric (\ref{g3}) into Boyer-Lindquist coordinates.  For this, we use a global coordinate transformation \citep{Azreg-Ainou:2014pra} given by 
\begin{equation}
	du=dt'+\lambda(r)dr, \hspace{0.5cm} d\phi=d\phi'+\chi(r) dr,
	\label{trans}
\end{equation}
where $\lambda(r)$ and $\chi(r)$ are functions of $r$ only, which are calculated as \citep{Azreg-Ainou:2014pra}
\begin{equation}
	\lambda(r)=-\frac{r^2+a^2}{f(r)r^2+a^2},\;\;\; \chi(r)=-\frac{a}{f(r)r^2+a^2}.
\end{equation}
Setting the coefficient of cross-term $dtdr$ in the metric to zero, we obtain
\begin{equation}
	F(r,a,\theta)=\frac{f(r)r^2+a^2 \cos^2\theta}{H(r,a,\theta)},
\end{equation}
and for the vanishing Einstein tensor component, $G_{r \theta}=0$, we obtain $H(r,a,\cos\theta)=r^2+a^2 \cos^2\theta$ \citep{Azreg-Ainou:2014pra}. After some manipulation, the rotating black hole metric in Boyer-Lindquist coordinates can be finally expressed as  
\begin{align}\label{metric_KEGBB}
ds^{2}=&-\left[\frac{\Delta-a^{2}\sin^{2}\theta}{\Sigma}\right]dt^{2}+\frac{\Sigma}{\Delta}dr^{2}+\Sigma d\theta^{2}\nonumber\\
&-2a\sin^2{\theta}\left[1-\frac{\Delta-a^2\sin^2{\theta}}{\Sigma}\right]dtd\phi \nonumber\\
&+\frac{\sin^2{\theta} }{\Sigma}\left[\left(r^{2}+a^2\right)^2-\Delta a^2\sin^2{\theta} \right]d\phi^{2}, \\
\text{with}&\nonumber\\
\Delta=&a^{2}+f(r),\quad \Sigma=r^2+a^{2}\cos^{2}\theta.\label{Sigma_KEGBB}
\end{align}
Thus, we obtained the rotating counterpart of metric \eqref{metric_spherical_transformed}, via the non-complexification procedure,  i.e., the metric of the Kerr black holes in EGB-bumblebee gravity, which we shall refer to as the Kerr EGB-bumblebee (KEGBB) black hole for brevity.
It may be true that our solution \eqref{metric_KEGBB} may not satisfy the field equations, a common trait shared with other rotating solutions in modified gravity \cite{Johannsen:2011dh,Jusufi:2019caq,Bambi:2013ufa,Moffat:2014aja,Kumar:2020owy}, and they are likely to generate extra stresses.
Thus, we regard our rotating metric \eqref{metric_KEGBB} as an EGB-bumblebee gravity black hole solution of an appropriately chosen set of unknown field equations but different from the Eqs. \eqref{bar}.
We note that this methodology has successfully been applied to generate imperfect fluid rotating solutions in Boyer$-$Lindquist coordinates from spherically symmetric static black holes \cite{Johannsen:2011dh, Brahma:2020eos, Kumar:2022fqo, Afrin:2021wlj, Kumar:2020owy}.  
The method has found application in producing rotating metrics within the framework of loop quantum gravity \cite{Brahma:2020eos} and in the context of rotating metrics in 4D EGB gravity \cite{Kumar:2020owy}. We have thus derived a rotating spacetime that
captures key aspects of quadratic curvatures and the bumblebee field, which is important from a phenomenological perspective. The resulting rotating solution \eqref{metric_KEGBB} reduces to the Kerr black hole of general relativity (GR) and approaches rotating metrics of 4D EGB gravity asymptotically.
\begin{figure}
    \centering
    \includegraphics[scale=0.65]{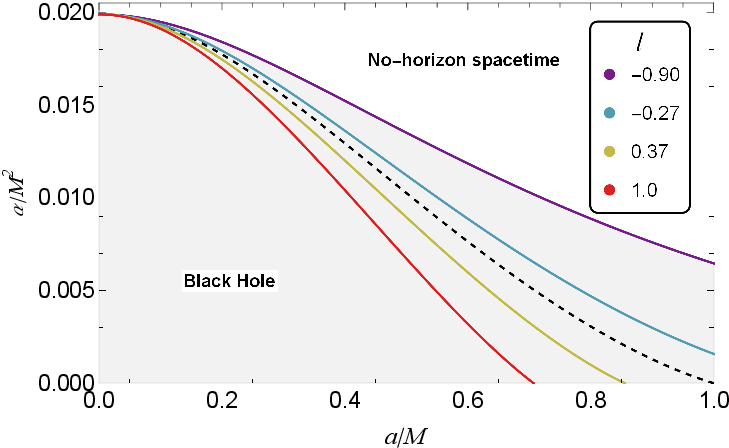}
    \caption{Spin parameter-GB parameter ($a$-$\alpha$) plane of the KEGBB black hole metric. The curves separate black holes from configurations without an event horizon or no-horizon spacetimes.  The black dashed curve corresponds to the $l=0$ case.}
    \label{parameterSpace}
\end{figure}
\begin{figure*}[t]
\centering
\begin{tabular}{c c}
    \includegraphics[scale=0.72]{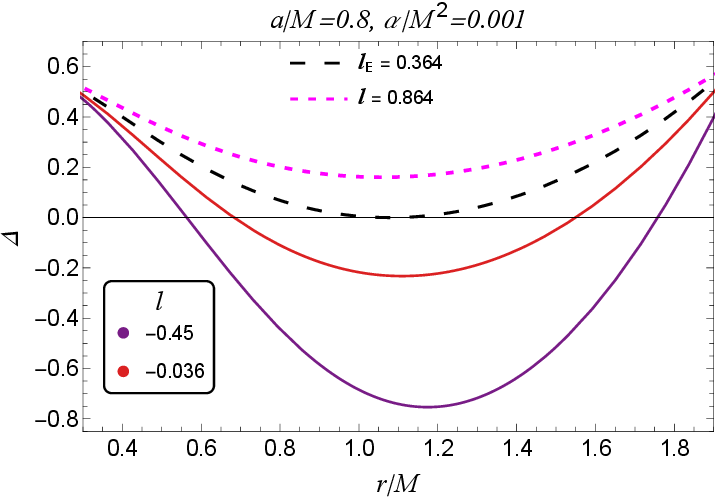}&
    \includegraphics[scale=0.72]{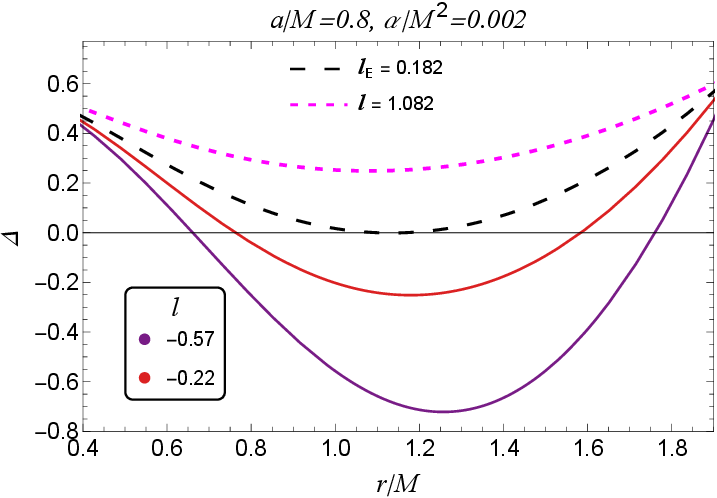}\\
    \includegraphics[scale=0.72]{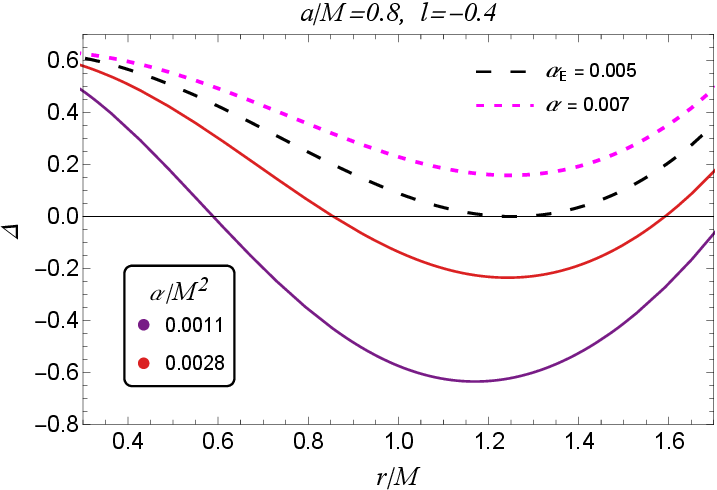}&
    \includegraphics[scale=0.72]{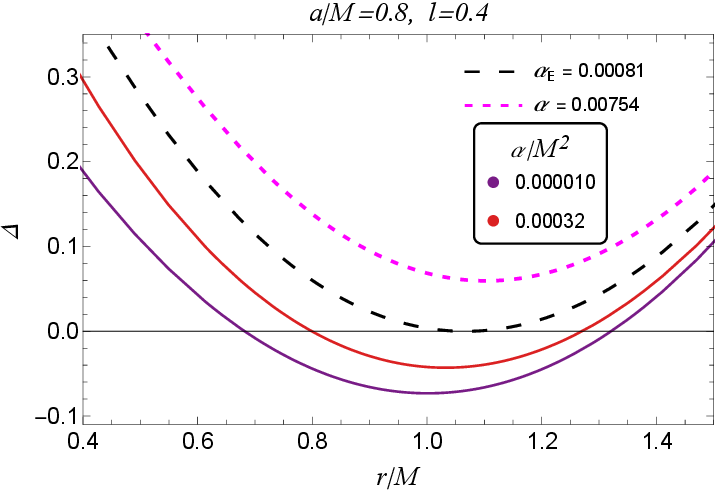}
\end{tabular}
\caption{\label{horizon_plot}Horizon structure of the KEGBB black holes with varying $l$ (top) and varying  $\alpha$ (bottom). The dashed black and magenta curves correspond, respectively, to an extremal black hole and no-horizon spacetime.}
\end{figure*}

In the KEGBB spacetime \eqref{metric_KEGBB}, a ring singularity occurs at $\Sigma=0$. Meanwhile, a coordinate singularity occurs when $\Sigma\neq0$ and $\Delta(r)=0$. Solving $g^{rr}=0=\Delta(r)$ we can obtain the horizon radii of the black holes \eqref{metric_KEGBB}. The condition for the existence of horizons for a given Lorentz violation parameter $l$ puts bounds on the black hole parameters $\alpha$ and $a$.
For a given set of parameters $a$, $l$ and $\alpha$, there are two possible solutions of $\Delta(r)=0$ which pertain to the two horizons  $r_{\pm}$; the event horizon being $r_{+}$. In the limit as $l\to0,~\alpha\to0$, these solutions converge to the horizons of the Kerr metric, which are given by $r_{\mp}^{K} = M\mp\sqrt{M^2-a^2}$. We illustrate the parameter space $(a, \alpha)$ for different Lorentz violation parameters $l$ in Fig.~\ref{parameterSpace}. Within the shaded region, a black hole with two horizons is found, and for all values of parameters ($\alpha_E$, $a_E$, $l_E$) on the solid lines, $r_\pm$ degenerate and an extremal black hole are obtained. For all parameters in the white region, no-horizon spacetimes are present. The allowed parameter space becomes constricted with increasing bumblebee parameter $l$.
In contrast to the spherically symmetric case, the bumblebee parameter $l$ profoundly impacts the horizons of the KEGBB black holes, as seen from Fig.~\ref{horizon_plot}. For given $\alpha$ and $a$, the Cauchy horizon radius increases, whereas the event horizon decreases with the increasing $l$. The extremal value $l_E$ is dependent on the $\alpha$ and $a$ for e.g., for $a=0.8M, \alpha=0.001M^2$, $l_E=0.364$ and for $a=0.8M, \alpha=0.002M^2$, $l_E=0.182$. The GB coupling parameter $\alpha$ similarly affects the $r_\pm$. The Cauchy and event horizon radii, respectively, increase and decrease with increasing $\alpha$. The extremal value $\alpha_E$ varies with $a$ and $l$, for e.g., $a=0.8M, l=-0.4$, $\alpha_E=0.005M^2$ whereas, for $a=0.8M, l=0.4$, $\alpha_E=0.00081M^2$.

Similar to the Kerr black holes, the KEGBB black holes possess time translational and rotational invariance isometries, as evinced by the existence of Killing vectors Killing vectors 
$\chi_{(t)}^{\mu}=\delta _t^{\mu }$  and $\chi_{(\phi)}^{\mu}=\delta _{\phi }^{\mu }$ respectively.  Due to the Lense-Thirring effect a static observer situated beyond the event horizon and possessing zero angular momentum relative to an observer located at spatial infinity will experience rotational motion along with the black hole's rotation \citep{Poisson:2009pwt}. In the KEGBB spacetime (\ref{metric_KEGBB}) the Lense-Thirring effect arises due to non-zero off-diagonal elements $g_{t\phi}$ and the angular velocity of the observer is given by \citep{Poisson:2009pwt},
\begin{eqnarray}\label{omega1}
    \tilde{\omega}=\frac{d\phi}{d t}=-\frac{g_{t\phi}}{g_{\phi\phi}}=\frac{a\left(a^2+r^2-\Delta\right)}{\left(r^2+a^2\right)^2-a^2 \Delta\sin^2{\theta}}.
\end{eqnarray}
The velocity $\tilde{\omega}$ exhibits a monotonic increase as the observer moves closer to the black hole, and on reaching $r=r_+$, the observer attains maximal co-rotation with a velocity equivalent to that of the black hole, which is given by
\begin{eqnarray}\label{omega2}
    \Omega= \tilde{\omega} |_{r=r_{+}}=\frac{a}{r_{+}^2+a^2}, 
\end{eqnarray}
which, in the limit $l\to0$, recovers the angular velocity of the KEGB black hole\citep{Kumar:2020owy} and further reduces to the angular velocity of Kerr black hole in the limit $\alpha\to0$ \citep{Chakraborty:2013naa}. Thus, the event horizon of the black hole exhibits a rigid body rotation \citep{Frolov:2014dta}, wherein every point on the $r=r_+$ surface has uniform angular velocity $\Omega$ when measured at infinity. 
\section{Black hole shadow}
The black hole, which, due to its immense gravity, lets not even light escape, cannot itself be seen. Still, its imprint on the surrounding light distribution can be seen as a dark region -- a shadow \citep{Islam:2022wck,KumarWalia:2022aop,Vagnozzi:2022moj,Ghosh:2021clx,Afrin:2021imp,Ghosh:2020spb,Kumar:2020yem,Kumar:2020ltt,Kumar:2020owy,Ghosh:2020syx,Kumar:2020hgm,Kumar:2019pjp,Kumar:2019ohr,Kumar:2018ple,Kumar:2017tdw}-- on the observer's sky whose outline is marked by gravitationally lensed photons. This phenomenon of gravitational lensing can be analytically explained based on the path a photon takes around the black hole. The photon geodesics, in turn, solely depend on the background spacetime's geometry and symmetries. Thus, obtaining the null geodesics would be directly informative regarding the MoG as well.
For the photon motion in spacetime \eqref{metric_KEGBB}, the four constants of motion: the rest mass $m_0=0$, energy $\mathcal{E}=-g_{t\phi}\dot{\phi}-g_{tt}\dot{t}$ and azimuthal component of angular momentum $\mathcal{L}_z=g_{\phi\phi}\dot{\phi}+g_{\phi t}\dot{t}$ along with the Carter constant $\mathcal{Q}$ arising from a fourth hidden spacetime symmetry, result in the first order null geodesic equations which can be obtained from the Hamilton Jacobi equation following Carter's separability approach \citep{Carter:1968rr, Chandrasekhar:1985kt}, given by
\begin{align}
\Sigma \dot{t}=&\frac{r^2+a^2}{\Delta}\left[\mathcal{E}\left(r^2+a^2\right)-a\mathcal{L}_z\right]\nonumber\\
&-a(a\mathcal{E}\sin^2{\theta}-\mathcal{L}_z),\label{teq}\\
\Sigma \dot{\phi}=&\frac{a}{\Delta}\left[\mathcal{E}\left(r^2+a^2\right)-a\mathcal{L}_z\right]-\left(a\mathcal{E}-\frac{\mathcal{L}_z}{\sin^2{\theta}}\right),\label{phieq}\\
\Sigma^2 \dot{r}^2=&\Big[\left(r^2+a^2\right)\mathcal{E}-a \mathcal{L}_z \Big]^2-\Delta \Big[{\mathcal{K}}+(a \mathcal{E}- \mathcal{L}_z)^2\Big]\nonumber\\
&\equiv\mathcal{R}(r)\ ,\label{req} \\
\Sigma^2 \dot{\theta}^2=&\mathcal{K}-\left(\frac{{\mathcal{L}_z}^2}{\sin^2\theta}-a^2 \mathcal{E}^2 \right)\cos^2\theta\equiv\Theta(\theta)\ .\label{theq}
\end{align}
Here the overdot represents differentiation to the affine parameter and $\mathcal{K}=\mathcal{Q}-(aE-L_z)^2$ is a separability constant \citep{Carter:1968rr,Chandrasekhar:1985kt}; the $\mathcal{R}(r)$ and $\Theta(\theta)$ are respectively the radial and angular potentials. The Equations~\eqref{teq}-\eqref{phieq} revert to KEGB black hole case for $l\to0$ \citep{Ding:2019mal} and further to the Kerr case when  $\alpha\to0$, $l=0$ \citep{Teo:2003}. Further, we introduce two energy re-scaled equations $ \xi=\mathcal{L}/\mathcal{E}$ and $\eta=\mathcal{K}/\mathcal{E}^{2}$
termed critical impact parameters in the $\mathcal{R}$ and $\Theta$ potential functions. We are interested only in the exterior photon region comprising the spherical photon orbits (SPOs), i.e., the null geodesics radially constrained to spherical photon orbit radii $r_p\,:\, \mathcal{R}'(r_p)=0=\mathcal{R}''(r_p)\,\forall\, r_p>r_+$ which yields the critical impact parameters around the KEGBB black holes,
\begin{align}
\xi_{c}=&\frac{\left(a^2+r^2\right) \Delta '(r)-4 r \Delta (r)}{a \Delta '(r)},\nonumber\\
\eta_{c}=&\frac{r^2 \left(8 \Delta (r) \left(2 a^2+r \Delta '(r)\right)-r^2 \Delta '(r)^2-16 \Delta (r)^2\right)}{a^2 \Delta '(r)^2}\label{CriImpPara},
\end{align}
and one recovers the critical impact parameters for Kerr spacetime for $l=0$, $\alpha\to0$ \citep{Chandrasekhar:1985kt}.
\begin{figure*}
\centering
\begin{tabular}{c c}
    \includegraphics[scale=0.82]{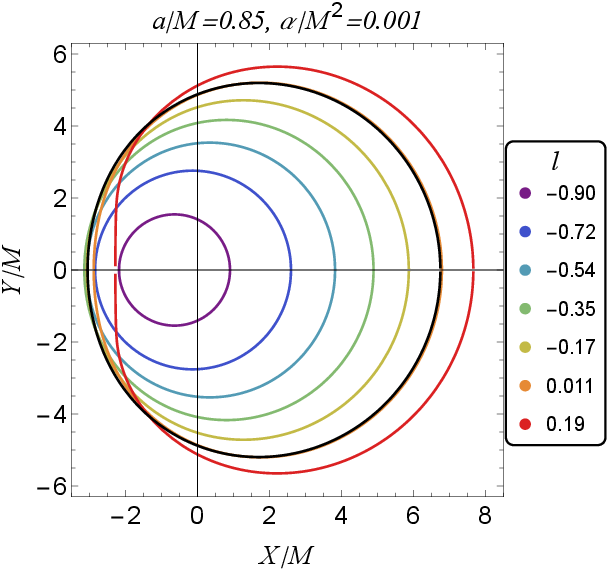}&
    \includegraphics[scale=0.82]{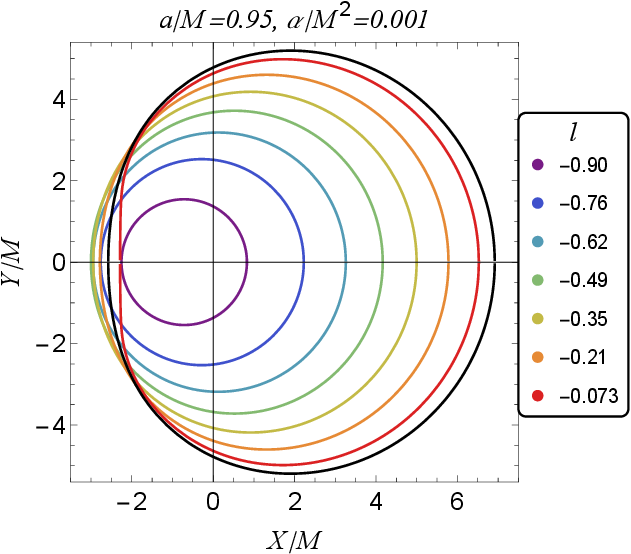}\\
    \includegraphics[scale=0.67]{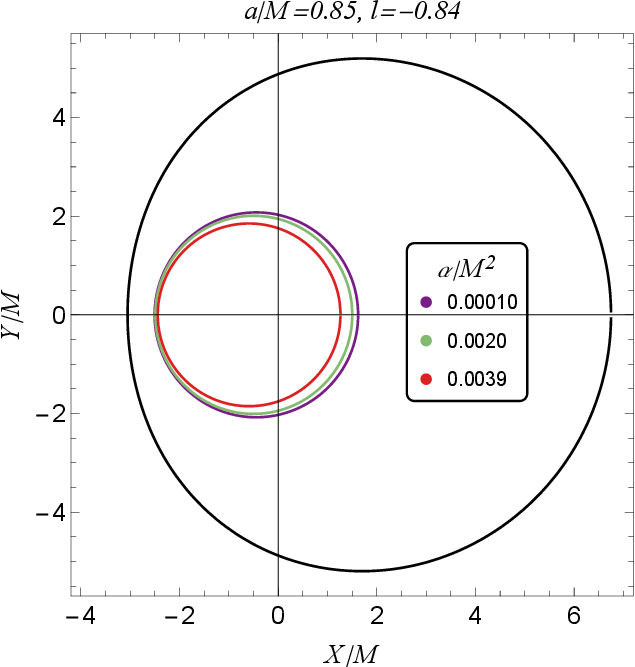}&
    \includegraphics[scale=0.67]{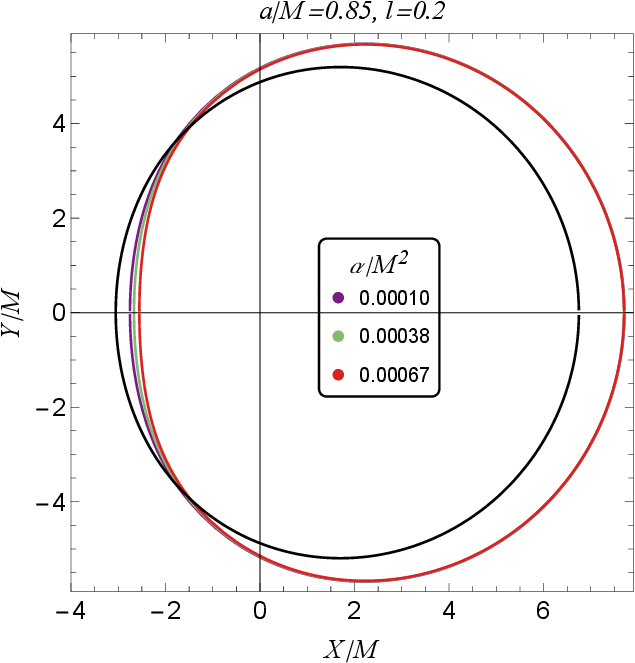}
\end{tabular}
\caption{Shadows of KEGBB black holes with varying $l$ (top) and $\alpha/M^2$ (bottom). The black curves are the shadows of the Kerr black holes.}\label{rotating_shadow_plot}
\end{figure*}

The SPOs are restricted to a 3-dimensional spacetime region 
--the photon shell \citep{Anjum:2023axh}--wherefrom, the photons marking the bright outline of the shadow silhouette originate \citep{Teo:2020sey,Johnson:2019ljv}. We can summarize the photon shell as,
\begin{eqnarray}
    t\in[-\infty, \infty],\;\;&	r_p\in[r_{p}^{-}, r_{p}^{+}],\;\; \phi\in[0, 2\pi], \nonumber\\
	\theta\in[\theta_{-}, \theta_{+}];\;\;&\theta_{\mp}\equiv\arccos\left(\mp\sqrt{\omega}\right),
\end{eqnarray}
where
\begin{equation}
    \omega=\frac{a^2-\eta_{c} -\xi_{c}^2+\sqrt{\left(a^2-\eta_{c} -\xi_{c}^2\right)^2+4 a^2 \eta_{c} }}{2 a^2}.
\end{equation}
Here $r_p^\mp$ are, respectively, the prograde and the retrograde photon radii given by the roots of $\eta_{c}=0$, such that $\xi_{c}(r_p^\mp)\gtrless0$, and the $\theta_{\mp}$ are the polar turning points obtained as the zeroes of $\Theta(\theta)=0$ \citep{Johnson:2019ljv}. 
\begin{figure*}
\begin{center}
    \begin{tabular}{c c}
    \includegraphics[scale=0.8]{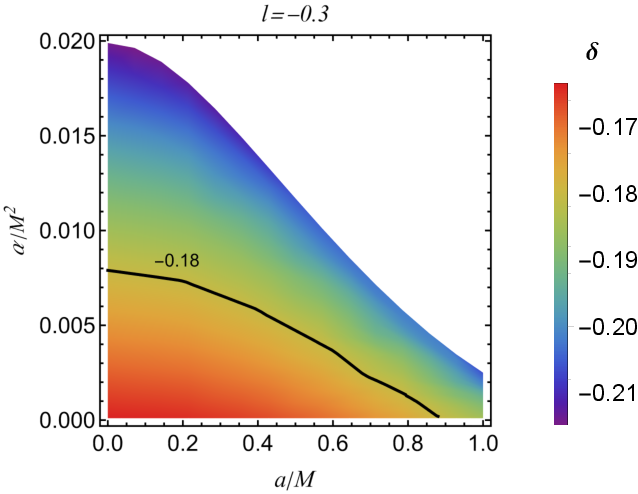}&
     \includegraphics[scale=0.8]{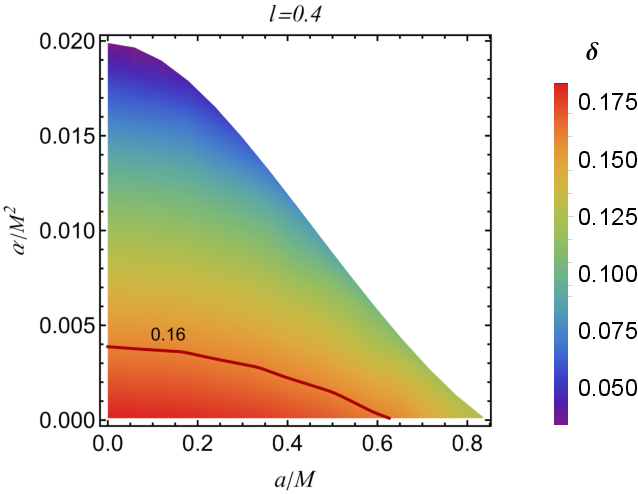}\\
     \includegraphics[scale=0.8]{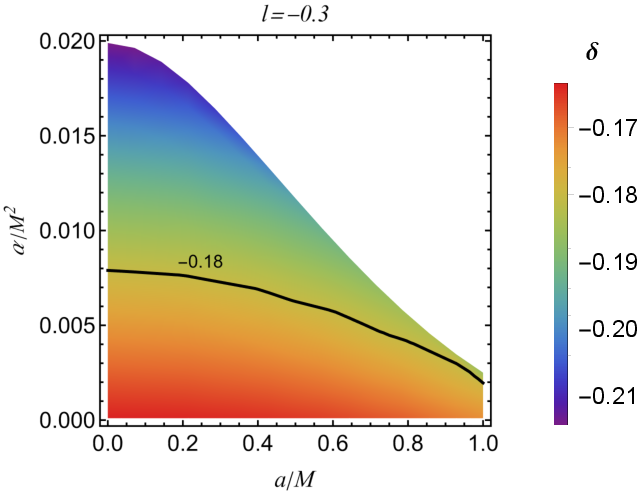}&
     \includegraphics[scale=0.8]{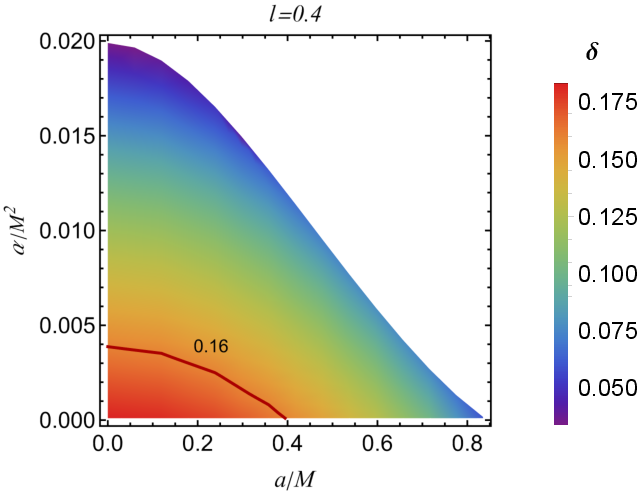}
\end{tabular}
\end{center}
	\caption{Plot showing Schwarzschild deviation $\delta$ of the KEGBB black hole shadows. It agrees with the EHT observations of the M87* black hole $\delta_{M87^*}=-0.01\pm0.17$. The inclination angle is $\theta_0=90$\textdegree (top) and $\theta_0=17$\textdegree (bottom). The white region pertains to no-horizon spacetime.}
	\label{Fig:delta_M87}
\end{figure*}
The photons transit from prograde to retrograde orbits at the intermediate value $r_p^0$ obtained by solving $\xi_{c}=0$; these are the SPOs with zero angular momenta \citep{Teo:2020sey}. At generic points in the photon shell, the SPOs oscillate in the $\theta$-direction between polar angles $\theta_{\mp}$ \citep{Johnson:2019ljv} but they are planar and confined to the $\theta=\pi/2$ plane at the boundaries of the annulus, $r=r_p^\mp$ \citep{Teo:2020sey,Afrin:2022ztr,Anjum:2023axh}. 

The SPOs form successive asymptotically spaced photon subrings that spiral-in to reach the bright boundary enclosing a dark region, namely the black hole shadow \citep{Johnson:2019ljv,Afrin:2021wlj,Anjum:2023axh}. Observationally, the shadow has been identified with the central flux depression outlined by an asymmetric bright ring-like structure as in the EHT captured images of SMBHs M87* and Sgr A* \citep{EventHorizonTelescope:2019dse,EventHorizonTelescope:2022exc}. Analytically, the shadow can be seen as the projection of a photon's shell onto an observer's celestial sky at spatial infinity.
\begin{figure*}
\begin{center}
    \begin{tabular}{c c}
    \includegraphics[scale=0.8]{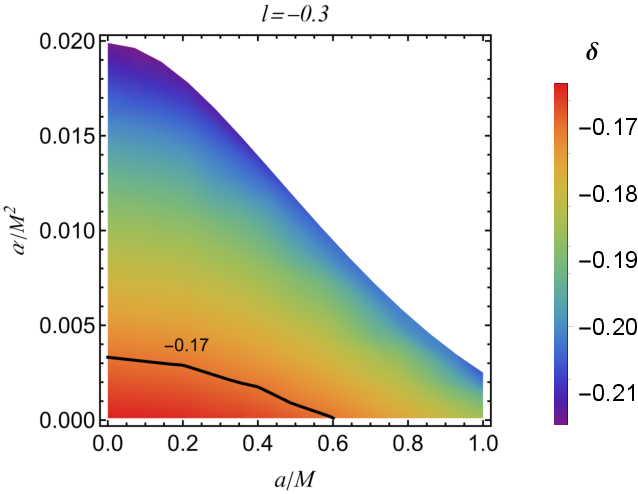}&
     \includegraphics[scale=0.8]{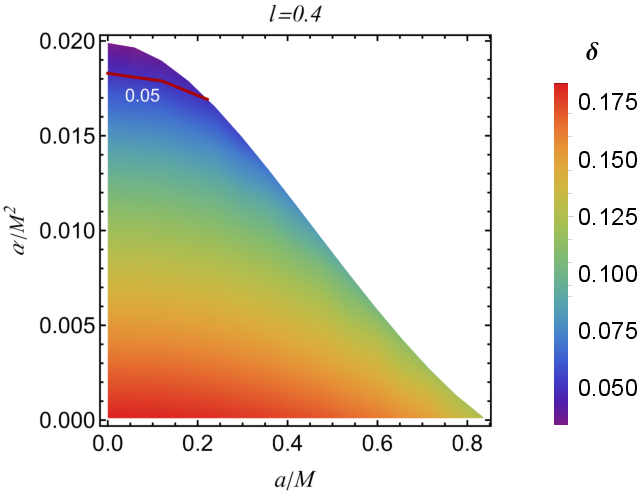}\\
     \includegraphics[scale=0.8]{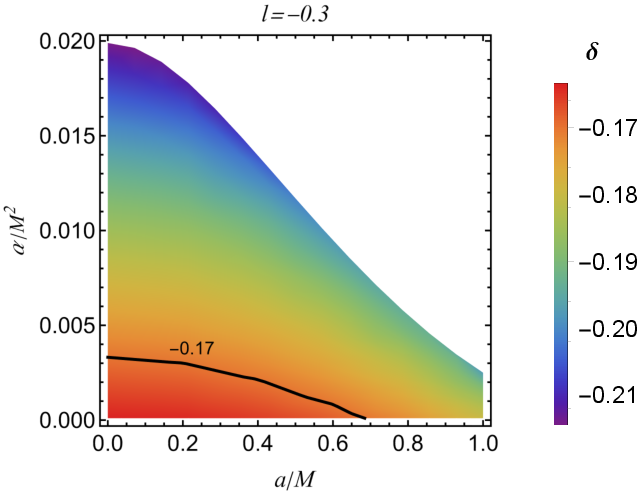}&
     \includegraphics[scale=0.8]{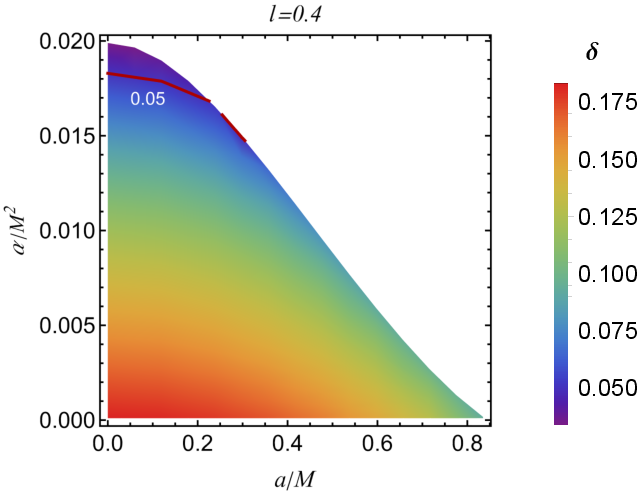}
\end{tabular}
\end{center}
	\caption{Plot showing Schwarzschild deviation $\delta$ of the KEGBB black hole shadows.  It is in agreement with the EHT observations of the Sgr A* black hole $\delta_{Sgr A^*}= -0.08^{+0.09}_{-0.09}~\text{(VLTI)},-0.04^{+0.09}_{-0.10}~\text{(Keck)}$. The black and red lines correspond respectively to the lower bound of VLTI and the upper bound of Keck. The inclination angle is $\theta_0=90$\textdegree (top) and $\theta_0=50$\textdegree (bottom). The white region pertains to no-horizon spacetime.}
	\label{Fig:delta_SgrA}
\end{figure*}
The shadow shape depends on the black hole parameters, i.e., spin and other hairs \citep{Johannsen:2015mdd,Afrin:2021imp,Afrin:2021ggx} alongside the observation angle $\theta_{0}$ relative to the spin axis, with the overall size scaled by the black hole mass $M$ \citep{EventHorizonTelescope:2019ths,Zakharov:2005ek}. Thus, for an observer at a distance $r_0\to\infty$ and an inclination angle $\theta_0$, the shadow would appear to be a dark region on the foreground of a bright background, outlined by the celestial coordinates, \citep{Bardeen:1973tla,Kumar:2018ple}
\begin{equation}
\{X,Y\}=\{-\xi_{c}\csc\theta_o,\, \pm\sqrt{\eta_{c}+a^2\cos^2\theta_o-\xi_{c}^2\cot^2\theta_o}\}\,\label{Celestial1}
\end{equation}
The shadow can be constructed by a parametric plot of ($X$, $Y$) in $r_p$.
The shadow cast by the KEGBB black holes is influenced by the $\alpha$ and $l$ parameters besides the spin $a$. Interestingly, for $l>0$, the shadows are bigger than the shadow cast by the Kerr black hole with same spin and are more distorted, whereas, for $l<0$, the shadows become smaller and less distorted; there is a horizontal shift of the shadow centre along the $x$-axis towards the retrograde side in the former case, and towards the prograde side in the latter case (see Fig.~\ref{rotating_shadow_plot}). This behaviour is similar to that induced by the spin $a$ in the case of Kerr black holes owing to the Lense-Thirring effect. However, the effect of the EGB coupling parameter $\alpha$ is not so profound, and the shadows only slightly vary with $\alpha$ (see Fig.~\ref{rotating_shadow_plot}). The degeneracy in the shadow characteristics of the KEGBB black holes and those of the Kerr black holes hint at the possibility of the former being suitable candidates for SMBHs, which are found to be described by the Kerr black holes in GR. We shall explore this possibility via observational results of the EHT experiment and go on to extract the black hole parameters from the shadow cast by KEGBB black holes.
\section{Testing KEGBB black holes as candidates for SMBHs with EHT results}
The EHT collaboration has released the images of SMBHs M87* \citep{EventHorizonTelescope:2019dse,EventHorizonTelescope:2019ggy} and Sgr A* \citep{EventHorizonTelescope:2022xnr,EventHorizonTelescope:2022xqj} which are highly accordant with the shadow cast by Kerr black holes and thus conform with the Kerr hypothesis \citep{Psaltis:2007cw}. However, due to significant uncertainties in the EHT results, MoGs cannot be ruled out. This opens an exciting possibility of probing the no-hair theorem \citep{Carter:1971zc} and, subsequently, a scope to test Lorentz violation in the strong field regime. The shadow characteristics are quantified by various observables describing the shadow shape and size. Besides, they encode the intricate details of the background spacetime. We shall employ the EHT bounds for the Schwarzschild deviation ($\delta$) of the shadow cast by the two SMBHs by modelling them as KEGBB black holes, to test the viability of the EGB-bumblebee gravity at the present observational resolution of the EHT. 
\begin{table}
\centering
\caption{Inferred EHT observational bounds on the parameters of SMBHs}
\label{Table:constraints}
\begin{tabular}{l|llll}
\hline
\multirow{3}{*}{SMBH} & \multicolumn{4}{l}{\quad\quad\quad\quad\quad\quad\quad\quad Upper limits}                    \\ \cline{2-5} 
                      & \multicolumn{2}{l|}{\quad\quad\quad\quad$l=-0.3$} & \multicolumn{2}{l}{\quad\quad\quad\quad$l=0.4$} \\ \cline{2-5} 
       & \multicolumn{1}{l|}{\quad$\alpha/M^2$} & \multicolumn{1}{l|}{\quad$a/M$}       & \multicolumn{1}{l|}{\quad$\alpha/M^2$} & \quad$a/M$        \\ \hline
M87*   & \multicolumn{1}{l|}{$[0, 0.008)$} & \multicolumn{1}{l|}{$[0, 0.88)$} & \multicolumn{1}{l|}{$[0, 0.004)$} & $[0, 0.39)$ \\ \hline
Sgr A* & \multicolumn{1}{l|}{$[0, 0.003)$} & \multicolumn{1}{l|}{$[0, 0.59)$} & \multicolumn{1}{l|}{$[0, 0.018)$} & $[0, 0.84)$  \\ \hline
\end{tabular}
\end{table}
\begin{figure*}
\begin{center}
    \begin{tabular}{c c}
     \includegraphics[scale=0.90]{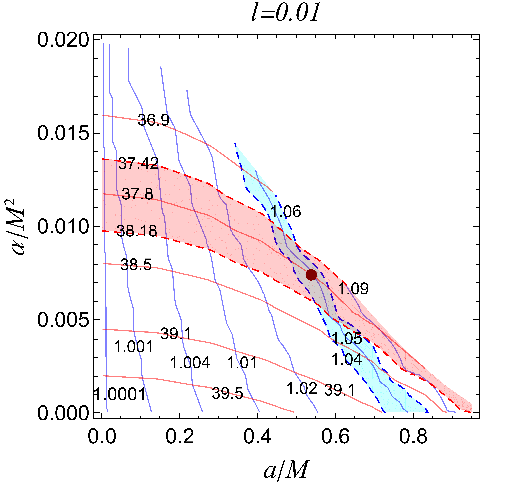}&
     \includegraphics[scale=0.90]{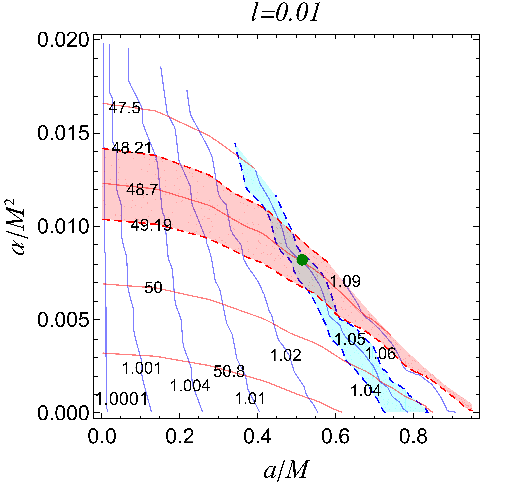}
\end{tabular}
\end{center}
	\caption{Contour plots of the EHT observables $d_{sh}$ and $\mathcal{D}_A$ in the plane $(a, \alpha)$ of KEGBB black holes. We label each curve with the corresponding values of $d_{sh}$ (red lines) and $\mathcal{D}_A$ (blue lines). The brown point (left) and the green point (right) denote the estimated parameters $\alpha$ and $a$ of SMBHs M87* and Sgr A*, respectively.}
	\label{Fig:Estimation}
\end{figure*}

To calculate the Schwarzschild deviation, we first obtain the shadow area, which is given by \citep{Kumar:2018ple,Afrin:2021imp}
\begin{equation}
A=2\int_{r_p^{-}}^{r_p^+}\left( Y(r_p) \frac{dX(r_p)}{dr_p}\right)dr_p,\label{Area}
\end{equation} 
The Schwarzschild deviation ($\delta$), which quantifies the difference between the model shadow diameter ($\Tilde{d}_{metric}$) and the Schwarzschild shadow diameter $6\sqrt{3}M$, is given by \citep{EventHorizonTelescope:2022xnr,EventHorizonTelescope:2022xqj},
\begin{equation}\label{SchwarzschildShadowDiameter}
    \delta=\frac{\Tilde{d}_{metric}}{6\sqrt{3}}-1,\,\, \Tilde{d}_{metric}=2R_a.
\end{equation}
Here $\Tilde{d}_{metric}$ is the model shadow diameter and $R_a=\sqrt{A/\pi}$ is the shadow's areal radius. 
As $a/M$ varies from 0 to $1$ and $\theta_0$ from 0 to $90$\textdegree, the shadow diameter of the Kerr black holes becomes smaller than the Schwarzschild shadow diameter by up to 7.5\%; thus $-0.075\leq\delta\leq0$  would result in the consistency of a model with Kerr whereas outside this range the model shadow would not conform with those cast by the Kerr black holes \citep{EventHorizonTelescope:2022xqj}. The KEGBB black holes cast shadows that are both smaller and larger than the corresponding Kerr shadows (see Fig.~\ref{rotating_shadow_plot}) for specific parameters ($\alpha, a, l$), thus, aided with the bounds on $\delta$ the KEGBB black holes can be tested as candidates for SMBHs. The first-ever image of SMBH M87*---obtained by the EHT using an extensive library of stimulated black hole images considering various combinations of the black hole parameters---shows an asymmetric bright ring produced by a combined effect of strong lensing and relativistic beaming and a central brightness depression, identified as the black hole shadow \citep{EventHorizonTelescope:2019dse}. The asymmetric ring has a diameter $(42\pm3)\mu$as which has been calibrated with the size of the black hole shadow to obtain the Schwarzschild deviation $\delta_{M87^*}=-0.01\pm0.17$ within $1\sigma$ confidence region \citep{EventHorizonTelescope:2019ggy,EventHorizonTelescope:2021dqv}.

The tests of the background MoG would, however, be accompanied by provisos during the analysis. There are various observational systematics that obfuscate the EHT observations that may arise because of the observations being made by different telescopes in a sparse array, the relatively uncertain astrophysics behind the radiation, plasma and accretion phenomena \citep{Gralla:2020pra,Gralla:2019xty}. Further, there is ambiguity in whether the ring-like feature seen in the images of both M87* and Sgr A* are indeed the gravitationally lensed images of the photon region or the surrounding accretion disk or a combination of both \citep{Gralla:2020pra,Gralla:2019xty}; if the latter is the case then future, more precise observations can detangle the astrophysical effects to resolve the photon rings which are dependent solely on the spacetime structure \citep{Ghosh:2022kit}. The EHT has, however, made calibrations to address many of the aforementioned uncertainties to obtain the bounds on the shadow observables. Hence, we consider the characteristics of the bright ring enclosing the brightness depression as the shadow and compute the shadow observables.

Additionally, when considering the position of the relativistic jet in the image of M87*, the EHT inferred the inclination angle to the line of sight to be approximately $163$\textdegree~ \citep{Walker:2018muw}. Since the current analysis does not take into account the accretion flow and only uses the analytical shadow curve, due to the top-bottom symmetric nature of the shadow, an inclination of $163$\textdegree~ $\sim$ $17$\textdegree~. However, the shadow exhibits maximum distortion only at very high inclinations, around $\theta_0\approx90$\textdegree. We shall conduct our analysis considering the observation angles $\theta_0=17$\textdegree~ and $90$\textdegree.

We take the inferred black hole mass $M_{M87^*}= 6.5\times10^9 M_\odot$ and distance $d_{M87^*}=16.8 Mpc$ from Earth to impose the EHT inferred bound on $\delta_{M87^*}$. From Fig.~\ref{Fig:delta_M87} we observe that a sizeable part of ($\alpha$, $a$) parameter space is consistent with the EHT results for M87* and we retrieve the bounds: (i) $\alpha/M^2\in[0, 0.008)$, $a/M\in[0, 0.88)$ at $90$\textdegree~, and $\alpha/M^2\in[0, 0.008)$, $a/M\in[0, 1)$ at $17$\textdegree~ for $l=-0.3$ and (ii)  $\alpha/M^2\in[0, 0.004)$, $a/M\in[0, 0.63)$ at $90$\textdegree~, and $\alpha/M^2\in[0, 0.004)$, $a/M\in[0, 0.39)$ at $17$\textdegree~ for $l=0.4$. Hence we infer that for the allowed range of parameters $\alpha/M^2\in[0, 0.008], a/M\in[0,0.88]$ for $l=-0.3$ and $\alpha/M^2\in[0, 0.004], a/M\in[0,0.39]$ for $l=0.4$, the KEGBB black holes can be viable candidates for SMBH M87* at the current precision of the astrophysical observations. Interestingly, the presence of Lorentz violation parameter $l$, significantly affects the upper bounds on $\alpha$ as for $l=0$ earlier studies find $\alpha/M^2\in[0, 0.00394)$ \citep{Kumar:2020owy}.

We consider next, the EHT results of SMBH Sgr A*, which opens a possibility of carrying out independent yet complimentary probes at a curvature scale $\sim\mathcal{O}(10^3)$ higher than probed with M87*. Unlike for M87*, for Sgr A*, the EHT has inferred both the shadow diameters and the diameter of the ring by a calibration technique taking into consideration various systematic and observational uncertainties and including different spacetime models in the imaging library \citep{EventHorizonTelescope:2022xqj}.  Further, independent priors are available for the mass-to-distance ratio of Sgr A*, with very small uncertainties, from stellar astrometry via two different experiments, namely the Keck and Very Large Telescope Interferometer (VLTI) ventures \citep{Do:2019txf,GRAVITY2019,GRAVITY2021,GRAVITY2022,EventHorizonTelescope:2022xqj}, which lowers the degree of freedom and gives a more accurate parameter-free prediction of the gravitational effects on the captured image of Sgr A* \citep{EventHorizonTelescope:2022xqj}. The EHT has put bounds on the Schwarzschild deviation for Sgr A*: $\delta_{Sgr A^*} = -0.08^{+0.09}_{-0.09}~\text{(VLTI)},-0.04^{+0.09}_{-0.10}~\text{(Keck)}$ at $1\sigma$ confidence level \citep{EventHorizonTelescope:2022xnr,EventHorizonTelescope:2022xqj}, that we intend to utilize. We will use the mass $M_{SgrA^*} = 4.0 \times 10^6 M_\odot $ and distance $d_{SgrA^*}=8 kpc$ priors  \citep{EventHorizonTelescope:2022xnr,EventHorizonTelescope:2022xqj} for calculating $\delta$.

Our analysis considers the observation angles $\theta_0=90$\textdegree~, $50$\textdegree~ and $0$\textdegree~ to test and constrain the EGB-bumblebee gravity. The EHT has disfavoured $\theta_0>50$\textdegree~  \citep{EventHorizonTelescope:2022xqj} for Sgr A*, but has obtained no estimates; thus, our analysis tackles one of the major sources of possible uncertainty in the EHT results for Sgr A* arising from the indetermination of the inclination, which is often lacking in theoretical investigations.
Imposing the EHT inferred bounds on $\delta$, and we set the bounds on the KEGBB black hole parameters: (i) $\alpha/M^2\in[0, 0.003)$, $a/M\in[0, 0.59)$ at $90$\textdegree~, $\alpha/M^2\in[0, 0.003)$, $a/M\in[0, 0.69)$ at $50$\textdegree~ and $\alpha/M^2\in[0, 0.003)$, $a/M\in[0, 0.94)$ at $0$\textdegree~ for $l=-0.3$, (ii)  $\alpha/M^2\in[0, 0.018)$, $a/M\in[0, 0.84)$ at all of the inclinations, i.e. $90$\textdegree~, $50$\textdegree~ and $0$\textdegree~, for $l=0.4$. Thus, we report that when $\alpha/M^2\in[0, 0.003], a/M\in[0,0.59]$ for $l=-0.3$ and $\alpha/M^2\in[0, 0.018], a/M\in[0,0.84]$ for $l=0.4$, the KEGBB black holes can describe the background spacetime of Sgr A*.
The good consistency of the parameter space with the observational results for both M87* and Sgr A* convey that the KEGBB black holes can be strong candidates for other SMBHs.

We emphasize our findings obtained by treating M87*  and Sgr A* as a  KEGBB black hole: Schwarzschild deviation $\delta$ of the KEGBB black hole shadows are in agreement with the EHT observations of the M87*  and Sgr A* black hole as shown in Figs. \ref{Fig:delta_M87} and \ref{Fig:delta_SgrA}; the obtained bounds on black hole parameters are tabulated in Table \ref{Table:constraints}. The remarkable consistency of the parameter space with the outcomes for both M87* and Sgr A* strongly suggests that KEGBB black holes could be robust contenders for SMBHs. By imposing the constraints derived from the EHT inferred bounds on the parameter $\delta$, and for chosen $l$, we can effectively restrict parameters $a$  and  $\alpha$  of KEGBB black holes.
 
\section{Black hole parameter estimation with EHT observables}
An important problem of theoretical and astrophysical importance is the extraction of the parameters of SMBHs. These include the intrinsic parameters, namely, the black hole's mass, spin and other charge parameters, and extrinsic ones, e.g., the black hole's distance, the inclination angle, etc. Though several methods exist to estimate the SMBH parameters, none alone is expected to estimate all the parameters. Only a concoction of various independent estimation techniques is expected to convey a complete description of the SMBHs. Afrin \& Ghosh \citep{Afrin:2023uzo}  have recently developed a novel technique using two shadow observables---the shadow angular diameter ($d_{sh}$) and the axial ratio ($\mathcal{D}_A$)---namely, the EHT observables to extract the spin, inclination and electric and magnetic charges of SMBHs. The method gives accurate estimates that agree with prior estimations from other techniques. We shall apply the method here to pinpoint the GB parameter $\alpha$ for a given bumblebee parameter $l$, which would, in turn, as a first,  give more information about dark energy directly through the shadow cast by the SMBHs.

\begin{table}
\centering
\caption{Estimated SMBH parameters using EHT observables for $l=0.01$.}
\label{Table:estimation}
\begin{tabular}{c|cc}
\hline
SMBH   & \quad Estimated parameters\hspace{-1cm}   &  \\ \cline{2-3} 
       & \multicolumn{1}{c|}{$\alpha/M^2$}              & $a/M$                  \\ \hline
M87*   & \multicolumn{1}{c|}{$0.007^{+0.003}_{-0.003}$} & $0.54^{+0.13}_{-0.11}$ \\ \hline
Sgr A* & \multicolumn{1}{c|}{$0.008^{+0.003}_{-0.003}$} & $0.51^{+0.10}_{-0.09}$ \\ \hline
\end{tabular}
\end{table}

We next model M87* and Sgr A* as  KEGBB black holes and calculate the angular diameter of the shadows cast using
\begin{align}
d_{sh}=2\frac{R_a}{d_{bh}},\label{angularDiameterEq}
\end{align}  
where $bh\equiv\{$M87*, Sgr A*$\}$, respectively, for the two SMBHs. 
The $d_{sh}$ encapsulates the scaling of the shadow size with the mass-to-distance ratio of the SMBHs. However, besides the size, the shadow shape must be considered to render an exhaustive description. The deviation from a circular shape gives one such measure, which may allude to a nonzero inclination angle, black hole spin and hair parameters. We quantify the circular asymmetry in the shadows cast by the SMBHs in terms of the axial ratio $\mathcal{D}_A$, which is the ratio of major to minor diameters of the shadow \citep{EventHorizonTelescope:2019dse}, and is given by \citep{Afrin:2021imp}
\begin{equation}
    \mathcal{D}_A=\frac{\Delta Y}{\Delta X},
\end{equation}
The ring diameter of M87* has been measured to be $(42 \pm 3)\mu$as \citep{EventHorizonTelescope:2019dse} and considering the $\lesssim10\%$ offset between the mean diameter of the ring and the photon ring diameter \citep{EventHorizonTelescope:2019ggy}, we calibrate the mean angular shadow diameter $d_{sh}^{M87^*}=37.8\mu$as. The EHT collaboration using calibration methods has obtained the shadow angular diameter of Sgr A* (47.8$\pm$7)$\mu$as 
\citep{EventHorizonTelescope:2022xnr}. We shall use the mean shadow diameter $d_{sh}^{Sgr A^*}=48.7\mu$as.
Further, the observed ring-like feature in the image of M87* is highly symmetric such that $1<\mathcal{D}_A\lesssim 1.33$ has been inferred by the EHT \citep{EventHorizonTelescope:2019dse,EventHorizonTelescope:2019ggy}, whereas, with the constraints on the ellipticity $0<\tau<0.5$, the axis ratio has been inferred to be $\sim 2:1$ \citep{Tiede:2022bdd}. However, due to sparse interferometric coverage, no information has been retrieved for the axis ratio of the shadow cast by Sgr A*. However, with future greater baseline coverage, we expect to have constraints on the $\mathcal{D}_A$ of Sgr A* soon. Besides, the bright emission rings exhibited by M87* and Sgr A* are comparable in their symmetric shapes. Further, for Kerr black holes $1\leq\mathcal{D}_A\leq1.1$ \citep{Tsupko:2017rdo}. Hence, in our analysis, we shall consider the axis ratio to be the mean value $\mathcal{D}_A=1.05$ for both M87* and Sgr A* at the present uncertainty of the EHT observations to demonstrate our method \citep{Afrin:2023uzo}. The directly measured values of the EHT observables have a $\sim14\%$ error bar owing to the observational, calibration and other uncertainties in measurement. However, future, more accurate observational runs, viz., that of the ngEHT might substantially reduce the error, enabling us to obtain accurate estimates of the SMBH parameters. We shall thus consider a $\leq1\%$  error bar on the presently measured mean values of the EHT observables, to see whether the SMBH parameters can be extracted from the shadow characteristics with our method and find the dependence of the estimated parameters on the measurement errors \citep{Afrin:2023uzo}.

From Fig.~\ref{rotating_shadow_plot}, we can surmise that there may exist some combinations of parameters ($\alpha$, $a$, $l$) for which the shadow cast by the KEGBB black holes are degenerate with those cast by the Kerr black holes with some spin parameter $a_*$, besides more that one combination of  ($\alpha$, $a$, $l$) may give rise to the same shadow cast by the KEGBB black holes. Hence, the shadow observables may degenerate in the parameters of the KEGBB black holes. We explore this possibility by constructing constant contours of  $d_{sh}$ and $\mathcal{D}_A$ in the parameter space of the KEGBB black holes. We find that the shadows of the KEGBB black holes degenerate for infinite parameter points on the contour of a given observable. However, the contours of the two observables intersect at unique points, and thus, we can extract any any two black hole parameters from the intersection point (see Fig.~\ref{Fig:Estimation}).
Since we can estimate any two parameters at once with our method, we fix the inclinations of both the SMBHs at $\theta_0=90$\textdegree. For M87*, we get the estimated values of the parameters to be $l=0.01$, $\alpha=0.007^{+0.003}_{-0.003}M^2$ and $a=0.54^{+0.13}_{-0.11}M$ (see the left panel of Fig.~\ref{Fig:Estimation}), whereas, the parameters of Sgr A* are estimated to be $l=0.01$, $\alpha=0.008^{+0.003}_{-0.003}M^2$ and $a=0.51^{+0.10}_{-0.09}M$ (see the right panel of Fig.~\ref{Fig:Estimation}), that we tabulate in Table~\ref{Table:estimation}. We can carry out the estimation for other values of parameter $l$. Thus, we obtain, as first, precise estimations of the GB coupling for a given bumblebee parameter of the two SMBHs M87* and Sgr A*, via EHT observables extending the earlier methods that yielded theoretical estimates of $\alpha$ \citep{Kumar:2020owy}. Also, unlike earlier estimation methods utilizing the black hole shadow, the estimated parameters with our method are dependent on the measurement uncertainties of the directly observed quantities and, hence, are closer to accuracy. Further, the estimations will likely be improved with better, more accurate observational runs of the present and future EHT experiments.

\section*{Conclusion}
The EGB gravity theory possesses several promising characteristics that set it apart from Einstein's general relativity and distinguish it from other higher-curvature theories. In 4D, the GB action is topological and does not influence gravitational dynamics. Nevertheless, by employing a consistent dimensional regularization approach, one can extract the non-trivial contribution of the GB term to the 4D gravitational field equations. The 4D regularized EGB gravity theories allow for spherically symmetric black holes characterized by two horizons depending on their critical mass. In contrast, the bumblebee gravity model extends the Einstein-Maxwell theory that permits spontaneous symmetry breaking, resulting in the field acquiring a vacuum expectation value, thereby introducing Lorentz violation into the system.
An exceptional opportunity to examine the strong-field predictions of GR and gain insights into metric theories of gravity arises from the observations made by the EHT for SMBHs M87* and Sgr A*. These observations have prompted us to explore Kerr black holes in a novel configuration: the 4D EGB bumblebee gravity model, where Lorentz symmetry is spontaneously broken.

Our primary goal is to evaluate how the background resulting from the spontaneous breaking of Lorentz symmetry and the presence of the Gauss-Bonnet parameter affect the properties of the Kerr black hole. Thus, we look for rotating black hole solutions within the framework of the EGB-bumblebee gravity model. This model action's unique feature lies in coupling the bumblebee vector field, the Ricci tensor, and the GB term. This intricate coupling enriches the theoretical framework, creating a spherical class of black hole solutions \eqref{metric_spherical}. Also, the limited availability of rotating black hole models in this context has impeded progress in subjecting them to observational tests. To address this challenge, we obtain the KEGBB solution, which have an additional Lorentz breaking parameter and GB parameter than the Kerr black hole, and they yield deviation from Kerr geometry but with a richer configuration of horizons and black hole shadow. The KEGBB permits studying the effect of higher curvature and Lorentz violation on the Kerr black holes and the result of Lorentz breaking parameter on rotating EGB black holes. We utilize the EHT shadow observations of M87* and Sgr A* to place constraints on the parameters of the KEGBB model.

Our investigation reveals that the Lorentz violation parameter $l$ profoundly influences the horizon's structure, reducing its radius. When considering fixed values of $\alpha$ and $a$, we observe that the Cauchy horizon radius increases as $l$ grows while the event horizon decreases. The extreme value $l_E$ is contingent on $\alpha$ and $a$. The radii of the horizons noticeably reduce as the parameters $\alpha$ and $a$ increase, resulting in changes to the ergosphere area and potentially intriguing consequences for the astrophysical Penrose process. 
The recent EHT observations of M87* and Sgr A* black hole shadows persuade us to explore the shadow cast by the KEGBB by examining the photon geodesics, which we could solve in the first-order differential form. Our analyses reveal that the KEGBB black holes cast shadows smaller and bigger than the Kerr black holes depending on the $l$ and $\alpha$. Notably, when $l > 0$, it's intriguing to observe that the shadows exhibit bigger sizes and are more distorted than those cast by Kerr black holes. Conversely, when $l < 0$, the shadows decrease in size with a decrease in $l$, and distortion also decreases. In the former scenario, the shadow's center shifts horizontally along the x-axis toward the retrograde side. In contrast, in the latter, it shifts towards the prograde side (refer to Fig.~\ref{rotating_shadow_plot}). However, the influence of the EGB coupling parameter $\alpha$ is comparatively moderate.

The black shadow observables that describe the shadow shape and size quantify the shadow characteristics. We employed the EHT observable Schwarzschild deviation ($\delta$) of the shadow cast by the SMBHs M87* and Sgr A* by modelling them as KEGBB black holes to test the viability of the 4D EGB-bumblebee gravity at the current observational resolution of the EHT. Our investigation reveals that the EHT observations of M87* and SgrA*, when considering a specific Lorentz violation parameter $l$, lead to alterations in the previously established upper limits for $\alpha$ and $a$. 
In conjunction with the EHT constraints for M87* and Sgr A*, our analysis indicates an agreement of a significant region of parameter space of KEGBB black holes with the EHT observations. It is possible, therefore, to assert that the KEGBB black holes could be reliable candidates for astrophysical black holes.
Finally, we employ our newly developed parameter estimation technique \citep{Afrin:2023uzo} wherein we apply two EHT observables, the shadow's angular diameter ($d_{sh}$) and the axial ratio ($\mathcal{D}_A$), to calculate the parameters of both M87* and Sgr A* considering the errors associated with the EHT observational data.

Our results constrain the black hole parameters and indicate that the EHT shadow observations do not entirely rule out KEGBB black holes. The departure produced by the KEGBB black hole parameters $l$  and $\alpha$ for the EHT results is $\mathcal{O}(\mu$as). 
Thus, from our perspective, distinguishing a KEGBB black hole from KEGB and Kerr black holes by observing black hole shadows is challenging. In the meantime, the forthcoming next-generation EHT (ngEHT) and space-based sub-millimetre interferometry technologies promise to deliver higher-resolution images of black holes. Fusing ngEHT imagery with gravitational wave observations may offer a more comprehensive framework for exploring theories of vector-tensor gravity similar to the Bumblebee model. We also expect significant constraints when considering the influence of the surrounding accretion disk. Furthermore, the generic equations of motion derived in terms of $\Delta(r)$ in this study would enable a detailed analysis of strong gravitational lensing. Besides, the Regge-Wheeler potential for calculating the quasinormal modes can also be studied. We intend to pursue both these investigations in a future work.

\section*{Acknowledgements}
 M.A. acknowledges DST-SERB for NPDF, file no. PDF/2023/000509, Department of Science and Technology, Govt. of India. S.G.G. thanks the SERB-DST, India for CRG project No. CRG/2021/005771.

\bibliography{bibfile}
\end{document}